\documentclass{iau}
\usepackage{graphicx}
\usepackage{amssymb}

\usepackage{wasysym}

\newlength{\digit}
\newcommand\tstrut{\rule{0pt}{2.5ex}}

\title[Collision parameters in early planetary systems] 
{Collision parameters governing water delivery and water loss in early
planetary systems}

\author[Thomas I.\ Maindl \& Rudolf Dvorak]   
{Thomas I.\ Maindl 
 \and Rudolf Dvorak}

\affiliation{
Universit\"atssternwarte, University of Vienna
\\ A-1180, Vienna, Austria
\\ email: {\tt thomas.maindl@univie.ac.at}
}

\pubyear{2013}
\volume{299}  
\pagerange{119--126}
\jname{Exploring the formation and evolution of planetary systems}
\editors{A.C. Editor, B.D. Editor \& C.E. Editor, eds.}
\begin{document}

\maketitle

\begin{abstract}
We investigate the distribution of encounter velocities and impact angles describing collisions in the habitable zone of the early planetary system.
Here we present a catalogue of collision characteristics for a particular mass ratio of the colliding bodies and seven different planetesimal masses ranging from a tenth of Ceres'
mass to 10 times the mass of the Moon.
We show that there are virtually no collisions with impact speeds lower than the surface escape velocity and a similar velocity-impact angle distribution for different planetesimal masses if velocities are normalized using the escape velocity.
An additional perturbing Jupiter-like object distorts the collision velocity and impact picture in the sense that grazing impacts at higher velocities are promoted if the perturber's orbit is close to the habitable zone whereas a more distant perturber has more the effect of a mere widening of the velocity dispersion.
\end{abstract}

\firstsection 
\section{Introduction}

\noindent The goal of this study is to quantify and characterize the number of collisions between
planetesimals in the early planetary system. Depending on the masses involved, we directly determine the velocity and the
direction of the collisions between these bodies. In this approach we do not yet intend to
accumulate them to larger ones; this is another task which is
actually already in the stage of computations. The next step---this is the core
of our overall project---is modeling the collisions in detail with our SPH code, which will allow us to track water content and explain water delivery processes by collisions and impacts (cf.\ Dvorak et al., 2012).

\subsection{The dynamical model and numerical setup}

\noindent As dynamical model we use full n-body integrations of the Sun and a ring of planetesimals in the habitable
zone. The bodies' initial orbital elements are uniformly distributed with $0.9<a<1.1\,\mathrm{AU}$ (semi-major axis), $e<0.1$ (eccentricity), and $i<1^{\circ}$ (inclination), respectively.
Additionally, we check the influence of a gas giant of $1\,\mathrm{M}_{\jupiter}$ in different distances
to the planetesimals which acts as a perturber in one particular scenario.

The scenarios differ in the chosen mass for the bodies - each run is done with a certain mass which varies from $5\cdot 10^{-11}\,\mathrm{M}_\odot\approx 1/10\,\mathrm{M}_\mathrm{Ceres}$ to $3\cdot 10^{-7}\,\mathrm{M}_\odot\approx 10\,\mathrm{M}_{\leftmoon}$. We use this relatively wide range of masses to see their
influence on the collision velocities and impact angles of the two bodies. Each scenario is integrated for $10^6$\,years and includes 750 planetesimals.
Table~\ref{t:nbodyscenarios} lists the considered n-body scenarios.

%
\begin{table}
\small
\begin{center}
\begin{tabular}{llclrr}
\hline
\hline
Scenario & $m\; [\mathrm{M}_\odot]$ & $m\; [\mathrm{kg}]$ & 
$R_\mathrm{imp}\; [\mathrm{10^6\,m}]$ & $r_\mathrm{Hill}\; [\mathrm{10^6\,m}]$ & $v_\mathrm{esc}\;[\mathrm{m}/\mathrm{s}]$\tstrut\\
\hline
Ce10 & $5\cdot 10^{-11}$ & $9.95\cdot 10^{19}$ & \hspace{2\digit}0.3918 &  38\hspace{2\digit} &  247\hspace{2\digit}\tstrut \\
Ce   & $5\cdot 10^{-10}$ & $9.95\cdot 10^{20}$ & \hspace{2\digit}0.8440 &  82\hspace{2\digit} &  532\hspace*{2\digit}\\
M10  & $3\cdot 10^{-9}$  & $5.97\cdot 10^{21}$ & \hspace{2\digit}1.534  & 150\hspace{2\digit} &  967\hspace*{2\digit}\\
M3   & $1\cdot 10^{-8}$  & $1.99\cdot 10^{22}$ & \hspace{2\digit}2.291  & 223\hspace{2\digit} & 1444\hspace*{2\digit}\\
M    & $3\cdot 10^{-8}$  & $5.97\cdot 10^{22}$ & \hspace{2\digit}3.304  & 322\hspace{2\digit} & 2083\hspace*{2\digit}\\
3M   & $1\cdot 10^{-7}$  & $1.99\cdot 10^{23}$ & \hspace{2\digit}4.936  & 481\hspace{2\digit} & 3112\hspace*{2\digit}\\
10M  & $3\cdot 10^{-7}$  & $5.97\cdot 10^{23}$ & \hspace{2\digit}7.119  & 694\hspace{2\digit} & 4488\hspace*{2\digit}\\
\hline
\end{tabular}
\end{center}
\caption{N-body simulation scenarios---$m\/$ is the planetesimal mass, $R_\mathrm{imp}$ denotes the mutual distance of the barycenters upon impact for $C_\mathrm{P}=0.3$, $C_\mathrm{T}=0$, $M_\mathrm{P}/M_\mathrm{tot}=0.1$, $r_\mathrm{Hill}$ the Hill radius at 1\,AU and zero eccentricity, and $v_\mathrm{esc}$ the target's surface escape velocity (see text).}
\label{t:nbodyscenarios}
\end{table}

\subsection{Collisions in an n-body context}

\noindent In point mass-based n-body simulations we assume a collision to happen if two objects experience a close encounter with mutual distance of their barycenters $R_\mathrm{imp}$ equal to the sum of their radii assuming spherical symmetry of the bodies. In each scenario this distance is significantly smaller than their Hill radius which---for an orbit with $a\approx 1\,\mathrm{AU}$ and small eccentricity---is $r_\mathrm{Hill}=\sqrt[3]{(m/(3\,\mathrm{M}_\odot)}\,\mathrm{AU}$ (cf.\ Tab.~\ref{t:nbodyscenarios}).

Hence, the collision can be treated as a two-body problem with a total mass of $M_\mathrm{tot}=2\,m$. For given mass $R_\mathrm{imp}$ will depend on the mass distribution between the impactors ($M_\mathrm{P}/M_\mathrm{tot}$ for the projectile, $M_\mathrm{T}/M_\mathrm{tot}$ for the target) and their radii $R_\mathrm{P}$ and $R_\mathrm{T}$ which in turn depend on their respective densities. As we are ultimately interested in water delivery by impacts we assume a solid basalt core with a layer or mantle of water ice, similar to the current models of Ceres (cf.\ Thomas et al.\ 2005).

Considering a body of mass $M\/$ consisting of basalt and a water ice shell (densities $\rho_\mathrm{b}$, $\rho_\mathrm{i}$) with a certain mass-fraction of ice $C\/$ we get the following relation for its radius:
\begin{equation}
R^3 = \left[C+(1-C)\,\frac{\rho_\mathrm{i}}{\rho_\mathrm{b}}\right]\,M\,\frac{3}{4\pi\,}\,\frac{1}{\rho_\mathrm{i}}.
\label{eq:R3}
\end{equation}
Keeping the future SPH-based simulations in mind we adopt the same values for the densities of ice and basalt as in Maindl et al.\ (2013): $\rho_\mathrm{i}=917\,\mathrm{kg}/\mathrm{m}^3$ and $\rho_\mathrm{b}=2,700\,\mathrm{kg}/\mathrm{m}^3$.

As a representative parameter set for collisions of interest for water delivery we choose a scenario with $M_\mathrm{P}/M_\mathrm{tot}=0.1$ and respective water contents $C_\mathrm{P}=30\,\%$ and $C_\mathrm{T}=0$.
\begin{figure}[b]
\begin{center}
{\includegraphics[width=0.28\textwidth, height=0.28\textwidth, angle=90]{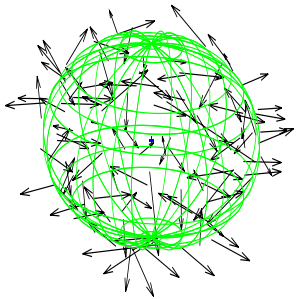}}
 \caption{Schematic view of impact velocities in the 3M scenario. The arrows show the negative impact velocity for better visibility. The majority of impacts occurs from low inclination orbits.}
   \label{f:vdirection}
\end{center}
\end{figure}

\section{Results}

\begin{figure}
\begin{center}     
{\includegraphics[width=0.91\textwidth]{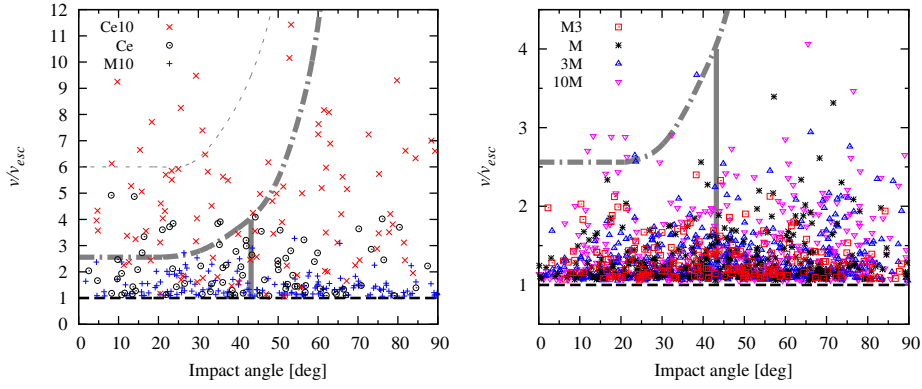}}
 \caption{Collision velocities normalized by the surface escape velocities versus impact angle for the different scenarios. Note that high velocities only occur in the Ce10 scenario. The lines refer to the boundaries of different collision outcomes as given in Leinhardt~\& Stewart (2012): net erosion to the target above the dot-dashed curve, below it partial accretion to the left and hit-and-run events to the right of the thick vertical line. There is no collision in the area of perfect merging (below the dashed line, $v/v_\mathrm{esc}\lesssim 1$). The dotted curve in the left diagram denotes the critical disruption velocity for half the total mass remaining in the largest remnant.}
\end{center}
   \label{f:vel}
\end{figure}

\noindent In each scenario multiple collisions occur. As the initial ``disk" of planetesimals is not considerably widened during the integration interval most of the impacts are close to the ecliptic (Fig.~\ref{f:vdirection}). As expected bigger masses correspond to bigger mutual perturbations and hence a larger number of collisions. The impact velocities show a relatively wide spread with a well defined lower boundary of about the escape velocity $v_\mathrm{esc}$ and an upper bound that varies between about 2.5 and 11.5 $v_\mathrm{esc}$. Figure~\ref{f:vel} shows scatter plots of velocities versus impact angles.

The results of all seven considered scenarios in absolute velocity units are summarized in Tab.~\ref{t:velocities}. The standard deviations along with the relatively large deviations of the mean and median values confirm the wide spread of impact velocities as observed in the scatter plots, especially for large-mass scenarios.
\begin{table}
\begin{center}
\begin{tabular}{lr@{\hspace{3\digit}}rrrr@{\hspace{3\digit}}rrrr@{\hspace{3\digit}}rrrr}
\hline
\hline
         & & \multicolumn{4}{c}{$0^\circ - 30^\circ$\hspace*{4\digit}} & \multicolumn{4}{c}{$30^\circ - 60^\circ$\hspace*{4\digit}} & \multicolumn{4}{c}{$60^\circ - 90^\circ$\hspace*{2\digit}}\tstrut \\
Scenario & $N_\mathrm{tot}$ & $N$ & $\bar{v}$ & $m_{v}$ & \multicolumn{1}{c}{$\sigma$\hspace*{3\digit}} & $N$ & $\bar{v}$ & $m_{v}$ & \multicolumn{1}{c}{$\sigma$\hspace*{3\digit}} & $N$ & $\bar{v}$ & $m_{v}$ & \multicolumn{1}{c}{$\sigma$} \\
\hline
Ce10 & 94 & 31 & 1.11 & 1.00 & 0.54 & 31 & 1.02 & 0.96 & 0.62 & 32 & 1.20 & 1.16 & 0.54\tstrut \\
Ce   & 79 & 23 & 1.22 & 1.03 & 0.64 & 41 & 1.16 & 1.15 & 0.47 & 15 & 1.07 & 0.74 & 0.57 \\
M10  &158 & 45 & 1.51 & 1.42 & 0.44 & 61 & 1.42 & 1.28 & 0.44 & 52 & 1.34 & 1.28 & 0.34 \\
M3   &195 & 62 & 1.96 & 1.83 & 0.41 & 91 & 1.92 & 1.76 & 0.38 & 42 & 1.87 & 1.84 & 0.29 \\
M    &303 & 80 & 2.65 & 2.50 & 0.47 &137 & 2.75 & 2.51 & 0.67 & 86 & 2.88 & 2.54 & 0.82 \\
3M   &449 &126 & 3.93 & 3.64 & 0.78 &197 & 4.11 & 3.84 & 0.94 &126 & 4.26 & 3.76 & 1.13 \\
10M  &483 &134 & 5.99 & 5.14 & 1.84 &201 & 6.03 & 5.40 & 1.56 &148 & 6.51 & 5.76 & 2.15 \\
\hline
\end{tabular}
\end{center}
\caption{Total number of collision events $N_\mathrm{tot}$, impact velocities (mean $\bar{v}$ and median $m_v\/$) and their standard deviations $\sigma$ for the different scenarios for three impact angle intervals. An impact angle of $0^\circ$ corresponds to a head-on collision. Velocity units are km/s.}
\label{t:velocities}
\end{table}

\begin{figure}
\begin{center}     
{\includegraphics[width=0.82\textwidth]{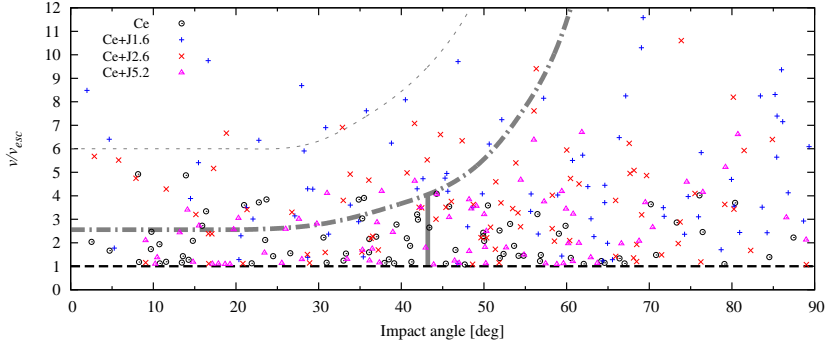}}
 \caption{The Ce scenario without and with a perturbing gas giant of $1\,\mathrm{M}_{\jupiter}$ placed at 1.6, 2.6, and 5.2\,AU from the Sun (denoted by Ce+J1.6, Ce+J2.6, and Ce+J5.2, respectively). The lines and symbols have the same meaning as in Fig.~\ref{f:vel}.}
   \label{f:velpert}
\end{center}
\end{figure}

Table~\ref{t:pert} and the plot in Fig.~\ref{f:velpert} show how different perturbing
``Jupiters'' effect the collision velocities in the Ce scenario. Especially a gas giant very close to the habitable zone significantly increases the spread in the velocities due to highly perturbed orbits. Also, hit-and-run collisions happen more often in that case. The significantly larger spread in the velocities may also increase the number of destructive collisions---an effect that decreases towards a ``real" Jupiter at larger distance $a_\mathrm{pert}=5.2\,\mathrm{AU}$.
\begin{table}
\begin{center}
\begin{tabular}{lrr@{\hspace{3\digit}}rrrr@{\hspace{3\digit}}rrrr@{\hspace{3\digit}}rrrr}
\hline
\hline
         & $a_\mathrm{pert}$ & & \multicolumn{4}{c}{$0^\circ - 30^\circ$\hspace*{4\digit}} & \multicolumn{4}{c}{$30^\circ - 60^\circ$\hspace*{4\digit}} & \multicolumn{4}{c}{$60^\circ - 90^\circ$\hspace*{2\digit}}\tstrut \\
Scenario & [AU] & $N_\mathrm{tot}$ & $N$ & $\bar{v}$ & $m_{v}$ & \multicolumn{1}{c}{$\sigma$\hspace*{3\digit}} & $N$ & $\bar{v}$ & $m_{v}$ & \multicolumn{1}{c}{$\sigma$\hspace*{3\digit}} & $N$ & $\bar{v}$ & $m_{v}$ & \multicolumn{1}{c}{$\sigma$} \\
\hline
Ce   & 1.6 & 80 & 17 & 2.50 & 2.28 & 1.38 & 27 & 2.62 & 2.23 & 1.46 & 36 & 2.71 & 2.27 & 1.60 \\
Ce   & 2.6 & 80 & 16 & 1.75 & 1.73 & 0.99 & 37 & 2.03 & 1.87 & 1.05 & 27 & 2.05 & 1.93 & 1.26 \\
Ce   & 5.2 & 71 & 19 & 1.03 & 0.84 & 0.45 & 35 & 1.39 & 1.30 & 0.74 & 17 & 1.68 & 1.41 & 0.94 \\
\hline
\end{tabular}
\end{center}
\caption{The effect of a perturbing body in the Ce scenario. The perturbing body's mass is $1\,\mathrm{M}_{\jupiter}$, its initial semi axis $a_\mathrm{pert}$. The other symbols and units are the same as in Tab.~\ref{t:velocities}.}
\label{t:pert}
\end{table}

\section{Conclusions and further research}

\noindent In our n-body calculations we confirm typical collision velocities in the early planetary system that range from the escape velocity $v_\mathrm{esc}$ up to a few times $v_\mathrm{esc}$ depending on the mass of the planetesimals. There is a clear tendency to collision velocities $v/v_\mathrm{esc}$ closer to 1 for heavier objects which may be explained by the two-body acceleration during the impact event dominating the initial velocity dispersion of the bodies for larger masses. The influence of perturbing bodies such as gas giants is significant in case of the perturber's orbit close to the planetesimals in the sense of a widened distribution of impact velocities and a tendency to promote hit-and-run collisions at high impact angles.

In the future, we will use collision velocities and angles corresponding to the partial accretion and hit-and-run scenarios in Figs.~\ref{f:vel} and~\ref{f:velpert} as input to analyzing water transport via collisions. We will use our 3D SPH code that---among other features---includes elasto-plastic material modeling, brittle failure, self-gravity, and first order consistency fixes. It is introduced in Sch\"afer (2005) and Maindl et al.\ (2013) and is still developed further.

\begin{acknowledgments}
\noindent The authors wish to thank C.\ Sch\"afer and R.\ Speith for jump-starting us on collisions of solid bodies and SPH as well as for numerous fruitful discussions. We also thank \'A.\ S\"uli and D.\ Bancelin for discussing and critically reviewing our n-body results.
This research is produced as part of the FWF Austrian Science Fund project S 11603-N16.
\end{acknowledgments}

\end{document}